\documentclass[a4paper,fleqn,usenatbib]{mnras}

\usepackage{newtxtext,newtxmath}

\usepackage[T1]{fontenc}
\usepackage{ae,aecompl}

\usepackage{graphicx}	
\usepackage{amsmath}	
\usepackage{amssymb}	

\title[Radio mode feedback: Does relativity matter?]{Radio mode feedback: Does relativity matter?}

\author[M. Perucho et al.]{
Manel Perucho,$^{1,2}$,\thanks{E-mail: manel.perucho@uv.es}
Jos\'e-Mar\'{\i}a Mart\'{\i},$^{1,2}$ Vicent Quilis,$^{1,2}$ Marina Borja-Lloret\,$^{1}$
\\
$^{1}$Departament d'Astronomia i Astrof\'{\i}sica, Universitat de
Val\`encia, C/ Dr. Moliner, 50, 46100, Burjassot, Val\`encia, Spain.\\ 
$^{2}$Observatori Astron\`omic, Universitat de Val\`encia, C/
Catedr\`atic Jos\'e Beltr\'an 2, 46980, Paterna, Val\`encia, Spain. 
}

\date{Accepted XXX. Received YYY; in original form ZZZ}

\pubyear{2015}

\begin{document}
\label{firstpage}
\pagerange{\pageref{firstpage}--\pageref{lastpage}}
\maketitle

\begin{abstract}
Radio mode feedback, associated with the propagation of powerful
outflows in active galaxies, is a crucial ingredient in galaxy evolution. 
Extragalactic jets are well collimated and relativistic, both in terms of
thermodynamics and kinematics. They generate strong shocks in
the ambient medium, associated with observed hotspots, and carve
cavities that are filled with the shocked jet flow. In this Letter, we
compare the pressure evolution in
the hotspot and the cavity generated by relativistic and classical
jets. Our results show that the classical approach
underestimates the cavity pressure by a factor $\geq 2$ for a given
shocked volume during the whole active phase. The tension between both
approaches can only be alleviated by unrealistic jet flow densities or
gigantic jet areas in the classical case. As a consequence, the
efficiency of a relativistic jet heating the ambient is typically
$\sim 20\%$ larger compared with a classical jet, and the heated
volume is 2 to 10 times larger during the time evolution. This
conflict translates into two substantially disparate manners, both
spatially and temporal, of heating the ambient medium.  These
differences are expected to have relevant implications on the star
formation rates of the host galaxies and  their evolution.  
\end{abstract}




\section{Introduction}

 Radio mode feedback is associated with the evolution of relativistic
jets in active galaxies. The way the energy deposited in the
ambient medium by these jets contributes to stop cooling flows
temporarily and quenches star formation within the host galaxy has
been addressed so far from two different assumptions: 1) the cavities triggered
by jets are in pressure equilibrium with their environments
\citep[see, e.g.][and references therein]{mn07,fb12}, and 2) the feedback is mediated 
by a shock \citep[as suggested by different numerical simulations][]{za05,wbu12,pe11,pe14}. The
results obtained from both points of view are obviously divergent: the
former studies typically conclude that heating might take place due to
mixing of the hot injected gas with the environment in the wake of a
buoyant, hot, and dilute bubble in the pressure gradient of the galaxy
or cluster; the latter conclude that the shock driving the cavity
expansion plays a fundamental role in heating, transferring a large
amount of the injected energy to the shocked ambient gas.
 
  Radio galaxies are classified in terms of their morphology, as FRI or
FRII. FRIs are known to be less powerful \citep{rs91,gc01}, with
a break power between $10^{44}$ and $10^{45}\,{\rm erg/s}$. Although
it is obvious that jets in FRII sources are surely triggering a
bow-shock \citep[e.g.,][]{cro11,sta14}, both observations
\citep{cro04,kr07} and numerical simulations \citep{pm07} have shown
that young FRI jets can also be surrounded by such shocks, albeit with
smaller Mach numbers than those in FRII jets.  
   
  Concerning the evolution of radio sources surrounded by shocks, a
strong debate still remains. Recent simulations have shown that the
efficiency of the energy transfer to the ambient medium, as well as
the spatial and temporal scales involved, can be very different for
relativistic (both kinematically and thermodynamically) and
non-relativistic jets \citep[][PMQR14 from now on]{pe11,pe14}. At very
large spatial and temporal scales, the total amount of energy released
to the ambient has to be the same in both, relativistic and
non-relativistic jets -- as it cannot be expected otherwise based on
conservation principles. However, the notable differences during the
jet evolution translate into relevant changes in the gas cooling rates
with time and position \citep[see, e.g.,][]{za05,oj10,gas12,cie14,guo16}.   
This conflict in the budget of cold gas fueling the galaxies could
lead to dramatic differences in the star formation rates in the host
galaxies and, therefore, in the galactic evolution as a whole.  

 In \citet{kf98}, the authors studied the differences on
the long-term evolution of relativistic and classical jets and found
an essential similarity in the evolution of both types of flow for jets
of the same power. Being in broad agreement with their conclusions, in this
Letter, we try to quantify the differences on the evolution of the
pressure at the hotspot and cavity in terms of the specific values
given to the parameters defining the jet power in both classical
and relativistic flows (internal energy, density, flow speed and jet
section). This is interesting to understand the process of energy
transfer to the ambient medium. Throughout this Letter, we will assume
an ideal gas equation of state to describe both the jet and the
ambient medium plasmas.  

\section{Relativistic versus classical approach}\label{mods}
   
  In this section we obtain analytical expressions for the comparison
of the hotspot pressure in relativistic and classical jets. In both
cases we assume that the terminal shock is strong, with both, shock
speed and post-shock flow speed much smaller than the initial jet
speed. 

\subsection{Classical jet} 
 
 The jet power for a classical jet is defined as:
\begin{equation}
 L_{j} = \left( \frac{\gamma_{j}\,P_{j}}{(\gamma_{j}-1)\,\rho_{j}}
   \,+\, \frac{1}{2} v_{j}^2 \right) \, \rho_{j}\, v_{j}\,A_{j}, 
\end{equation}
where $\gamma_j$ is the adiabatic exponent of the equation of state
describing the jet plasma, $P_j$ and $\rho_j$ stand for its pressure
and density, respectively, $v_j$ is the jet flow velocity, and $A_j$
is the jet cross-section. 

  In the case of a jet made of cold plasma, the jet power is then
simplified to the following expression: 
\begin{equation}\label{eq:L_c}
 L_{j} \simeq \frac{1}{2} v_{j}^3 \, \rho_{j}\,A_{j}.
\end{equation}

  The last relevant relation to be considered is the post-shock
pressure at the jet head for a strong shock, as derived from the
Rankine-Hugoniot jump conditions \citep{ll59,sch74,ka97}: 
\begin{equation}\label{p_c}
 P_{h} \simeq \frac{2\,\gamma_{j}\,P_{j}}{\gamma_{j} + 1}\,M_{j}^2, 
\end{equation}
where $P_h$ stands for the post-shock pressure at the
jet head, and $M_j$ is the jet Mach number. Comparing Eqs.~(\ref{eq:L_c})
and (\ref{p_c}) we obtain a relation between the jet  power and the
hotspot pressure: 
\begin{equation}\label{eq:p_Lc}
 P_{h} \simeq \frac{4\,L_{j}}{(\gamma_{j} + 1)\,v_{j}\,A_{j}}
\end{equation}

\subsection{Relativistic jet}        
  
   The total jet power for a relativistic jet is defined as:
\begin{equation}\label{eq:ljr}
 L_{j} = (h_{j}\,\Gamma_{j} - c^2) \, \Gamma_j\, \rho_{j} \, v_{j}\,A_{j}.
\end{equation}
where 
%
 $ h_{j} = c^2 + \frac{\gamma_{j} P_{j}}{(\gamma_{j} -1)\rho_{j}}$
%
is the jet specific enthalpy, $\Gamma_{j}$ is the jet Lorentz factor, and
$c$ is the speed of light. 

If the jet is hot ($h_{j} \gg c^2$) and has
a large Lorentz factor ($\Gamma_{j} \gg 1$), then:
\begin{equation}\label{eq:L_rel}
 L_{j}  \simeq \frac{\gamma_{j} P_{j}}{\gamma_{j}-1}\, \Gamma_{j}^2\, v_{j}\,A_{j}.
\end{equation}

 The conservation of momentum flux across
the shock is \citep[e.g.,][]{ll59,mm94}: 
\begin{equation}
 \frac{\rho_{j} \, h_{j} \, \Gamma_{j}^2 \, v_{j}^2}{c^2}+\, P_{j}\, = 
 \, \frac{\rho_{h} \, h_{h} \, \Gamma_{h}^2 \, v_{h}^2}{c^2} + \, P_{h}.
\end{equation}
Now, considering that the jet is both relativistic thermodynamically,
$h_{j} \gg c^2$ (i.e.,  $P_{j}/\rho_j \gg c^2$), and kinematically
($\Gamma_{j}^2 \gg 1$), and taking into account that if the shock is strong
$v_h \ll v_j$, the previous expression becomes
\begin{equation}\label{eq:p_rel}
P_{h}\simeq \frac{\gamma_{j} \Gamma_{j}^2 P_{j}}{\gamma_{j}-1} \, \frac{v_{j}^2}{c^2}.
\end{equation}
 Comparing Eqs.~(\ref{eq:L_rel}) and (\ref{eq:p_rel}), we obtain:
\begin{equation}\label{eq:p_Lrel}
 P_{h} \simeq \frac{L_{j} \, v_{j}}{A_{j} \, c^2}.
\end{equation}

Let us note that the last expression is also recovered in the case of
a cold relativistic jet as far as $h_j \Gamma_j \gg c^2$.
 
\subsection{Comparison between the classical and the relativistic approach}
  For the comparison between the classical and the relativistic
approaches, we include subscripts $c$ and $r$, respectively. 

\subsubsection{Jet power and jet cross-section}

  Since we are interested in comparing jets with the same power, 
according to Eqs.~(\ref{eq:L_c}) and (\ref{eq:ljr}), the
parameters defining both models shall verify
\begin{equation}
\frac{1}{2} v_{j,c}^3 \, \rho_{j,c}\,A_{j,c} \simeq h_{j,r}\,\Gamma_{j,r}^2\,
\rho_{j,r} \, v_{j,r}\,A_{j,r},
\end{equation}
or, equivalently, 
\begin{equation}\label{eq:cross}
\frac{A_{j,c}}{A_{j,r}} \simeq 74
 \left(
  \frac{v_{j,r}}{c}\right) \left(
\frac{v_{j,c}}{0.3c}\right)^{-3} \left(\frac{\rho_{j,r}}{\rho_{j,c}}\right) \left(\frac{h_{j,r}\,\Gamma_{j,r}^2}{c^2}\right).
\end{equation}

  A number of comments about this expression are relevant for our
discussion: 1) $v_{j,r} \simeq c$, 2) a jet velocity of $\approx
0.3c$ is in the limit of consistency with the non-relativistic
dynamics as the relativistic effects in the conservation of mass,
momentum and energy are in the range $5-10\,\%$, 3) we assume
$\rho_{j,c} \approx \rho_{j,r}$. Therefore, the first three terms in
parentheses are of order unity. This, together with the fact that the
last term can be much larger than one in the case of a hot,
relativistic jet, leads to the well known result that a classical jet
needs a section at least two orders of magnitude larger than a
relativistic jet of the same power.

\subsubsection{Hotspot pressure}

  The difference in the jet cross-sections of classical and
relativistic jet models of the same power has implications in the
ratio of hotspot pressures. From Eqs.~\ref{eq:p_Lc} and \ref{eq:p_Lrel}
we have
\begin{equation}
\frac{P_{h,c}}{P_{h,r}} \simeq \frac{4}{(\gamma_{j,c}+1)}
\frac{c^2}{v_{j,c} v_{j,r}} \frac{A_{j,r}}{A_{j,c}},
\end{equation}
or, substituting the cross section ratio from Eq.~\ref{eq:cross},
\begin{equation}
\frac{P_{h,c}}{P_{h,r}} \simeq 7 \times 10^{-2} \,
\left(\frac{\rho_{j,c}}{\rho_{j,r}}\right) \left(
\frac{c}{v_{j,r}}\right)^2 \left(\frac{0.3c}{v_{j,c}}\right)^{-2}
\left(\frac{c^2}{h_{j,r}\,\Gamma_{j,r}^2}\right)
\label{eq:pcpr}
\end{equation}
(where we have used $\gamma_{j,c}=5/3$, as corresponding to a cold
jet). Again the fact that the first three terms in parentheses are of
order unity and the last one can be significantly smaller than one,
leads to hotspot pressures of classical jets much smaller than those
of relativistic jets of the same power. If the relativistic jet 
    power was recovered by only increasing the jet rest-mass density,
    a classical jet would need $\rho_{j,c}\sim 10^4 \rho_{j,r}$ to
    obtain the same post-shock pressure as a mildly relativistic jet
    with $h_{j,r}\simeq 2c^2-10c^2$, $\Gamma_{j,r}\simeq5-10$.

\section{Results}\label{disc}

\subsection{Cavity pressure}
\label{ss:cp}
  The transfer of energy to the ambient medium in powerful radio
sources is driven by a strong shock that heats and accelerates the
ambient gas as it advances. The strength of this shock depends on the
evolution of the pressure in the cavity, $P_{cav}$, along time. Owing
to the large value of the sound speed in this region, this pressure is
very similar throughout the whole shocked volume (e.g., PMQR14).

  Since the cavity is inflated with expanding plasma flowing from the
hotspot, it is reasonable to expect that both cavity and hotspot
pressures are linked. The model presented in
  \cite{ka97} for pressure confined jets leads to $P_{cav} \propto
  P_{h}$, and a self-similar evolution of the source expansion
  at large times. The model was confirmed by numerical simulations of
classical and relativistic cold jets propagating in a uniform ambient
medium by \cite{kf98}. However it is expected that effects coming from
the internal energy content of the jets, or the propagation through
stratified media result in a more complex evolution. Indeed, the
hotspot evolution depends on the dynamics at the jet head and, in
particular, on the changes of the jet section close to the head. In
any case, we do expect two effects.  First, that the link between the hotspot and the cavity
remains. Second, that the systematic difference found analytically at early
stages between the hotspot pressures in both, the classical and
relativistic approaches, produces qualitative differences 
in the process of expansion of the cavity and, in particular, in the
efficiency of the heating of the ISM and IGM.

\subsection{Comparison with numerical simulations}

One way to estimate the efficiency of the heating is by comparing the
strength of the shocks in both classical and relativistic models
with the maximum shock strength achievable for a jet of a given
power which in turn is related to the quantity 
\begin{equation}
\label{eq:PV}
P_{cav,m} = \frac{L_{j}\, t}{V_{cav}},
\end{equation}
where $t$ is the age of the jet and $V_{cav}$, the cavity volume.
This expression represents the (maximum) pressure in the cavity
assuming that all the injected energy is transferred to it and participates in
driving the shock \citep{bc89}. 

\begin{table}
	\centering
	\caption{Parameters of the jets. First column: model name used in the text;
	 second column: jet velocity at injection in the grid; third column: jet rest-mass density; fourth column: jet specific enthalpy, and fifth column: jet radius at injection. The injection of the jets in the grid is set up at 1~kpc from the active nucleus.}
	\label{tab:xets}
	\begin{tabular}{lcccc} 
		\hline
		Model & $v_j$ ($c$) & $\rho_j$ ($g/cm^3$) & $h_j$ ($c^2$)& $R_j$ (pc)\\
		\hline
		J46c & 0.984 & $8.3\times10^{-29}$ & $1.0$ & 100\\
		J46h & 0.984 & $8.3\times10^{-30}$ & $8.0$& 100 \\
		J46n & 0.3 & $8.3\times10^{-29}$ & $1.0$ & $3\times10^3$\\
		\hline
	\end{tabular}
\end{table}

In the following we compare the time evolution of a set
of numerical simulations of classical and relativistic jet models
of the same power, paying attention to $i)$ the ratio
$P_{cav}/P_{cav,m}$, $ii)$ the volume swept by the shock in the
different simulations, and $iii)$ the efficiency in the energy
transfer to the ambient medium. 

The set of models is composed by a classical jet (J46n), a
relativistic cold jet (J46c), and a relativistic hot jet (J46h).  All
with a power of $10^{46}$ erg/s. Models J46c and J46n were already
discussed in PMQR14 (J46c being J46 in that work), and
J46h is new. We display the relevant parameters of the jets in Table~\ref{tab:xets}.
The rest-mass density of models J46c and J46n is the
same, but the jet radius of J46n was increased. The rest-mass density of the hot model was
reduced by a factor of 10 to obtain an equivalent jet power. Both relativistic jets have a Lorentz factor $\Gamma_{j}=5.6$. 

Models J46c and J46h propagate at an almost constant speed up to $t
\simeq 2$ Myr when they suffer an abrupt deceleration indicating the
transition from the one-dimensional (1D) propagation phase to the
two-dimensional (2D) propagation phase \citep[see, e.g.,][]{sch02,pe11}.
On the contrary, model J46n stays in the 1D phase along
the whole simulation. During the 1D phase of the
relativistic models, the ratio $P_{h,c}/P_{h,r}$ is much smaller than
one ($\simeq 0.08$), as 
anticipated by Eq.~(\ref{eq:pcpr}). Beyond this phase, the hotspot
pressures of the relativistic jets drop below the pressure
corresponding to the classical model and decrease steadily along the
evolution. The behaviour of the hotspot pressure is associated with
the change in the jet cross-section at the hotspot (see
Eqs.~\ref{eq:p_Lc} and \ref{eq:p_Lrel}, and Fig.~5 in PMQR14) which in
turn depends on the conversion of kinetic energy to internal energy
along the jet (much larger in the relativistic models).

  Regarding the relation between $P_h$ and $P_{cav}$,
we have applied $\upchi$-squared fits to the values obtained for each
simulation and found that $P_{cav} \propto P_{h}^{\, \alpha}$, with
$\alpha$ in the range $0.35$-$0.55$, depending on the simulation. The
small dispersion in the exponent of the power law indicates that the
physical processes linking the evolution of the hotspot and cavity
pressure are essentially the same for both classical and relativistic jets.

\begin{figure}
\begin{center}
\includegraphics[width=0.94\columnwidth]{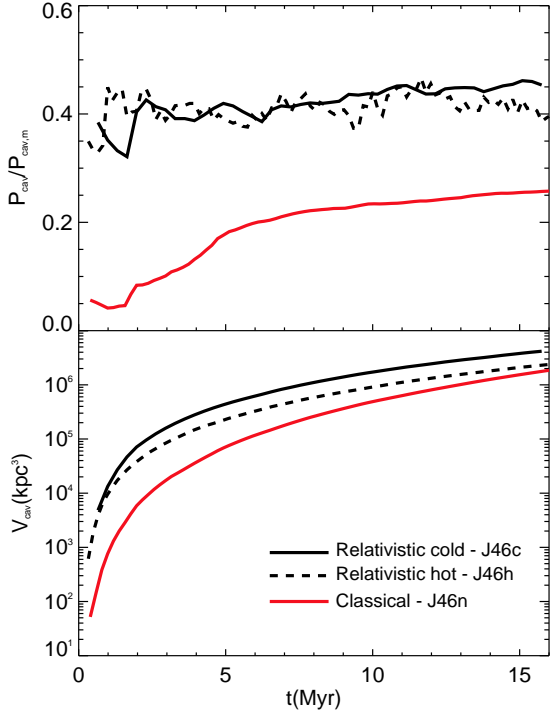}
\caption{Top panel: Ratio of the measured pressure at the cavity with
  respect to the maximum possible pressure, as given by
  Eq.~\ref{eq:PV}. Bottom panel: Shocked volumes. } 
  \label{fig:prat}
  \end{center}
\end{figure}

  Figure~\ref{fig:prat} (top panel) shows the evolution of the ratio
of the cavity pressure, $P_{cav}$, over the maximum cavity pressure
as defined by Eq.~(\ref{eq:PV}), for the three numerical
simulations. Both relativistic models show a similar ratio
$P_{cav}/P_{cav,m}$ ($\simeq 0.4$) that doubles the
corresponding to the classical model ($\leq 0.2$). The difference in
the pressure ratios starts right from 
the beginning extending through both the 1D and the
2D propagation phases of the relativistic jets. The
bottom panel of Fig.~\ref{fig:prat} shows the volume of the region
processed by the shock. The faster propagation of the relativistic
jets during their 1D phase and the drop of the hotspot pressure in the
relativistic models in the long term evolution translates into larger ambient
volumes affected by the shock (a factor 12 at the end of the 1D
phase; a factor 2 at $t = 16$ Myr).

%
%

\begin{figure}
\begin{center}
\includegraphics[width=0.94\columnwidth]{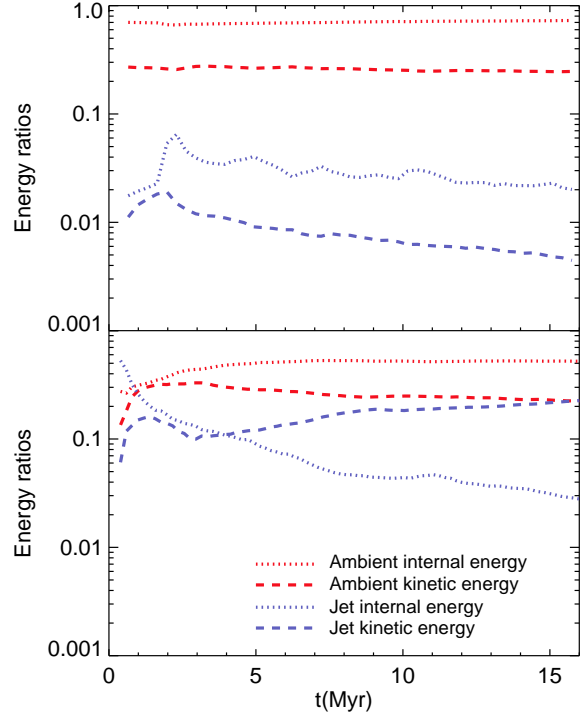}
\caption{Fraction of the injected energy invested in different
    channels as a function of time. The potential energy gained by
    the ambient gas is very small compared to those and is thus not
    plotted \citep[see][]{pe11}. Top panel: relativistic cold
    jet. Bottom panel: classical jet.} 
\label{fig:energy}
\end{center}
\end{figure}

  Figure~\ref{fig:energy} shows  the fraction of the injected energy
invested in different channels as a function of time for J46c and
J46n. The energy transfer efficiency to the ambient medium is higher
in the case of the relativistic jet. The difference in the amount of
kinetic energy kept by the jet particles in the two cases explains the
difference in $P_{cav}/P_{cav,m}$ between the two the classical and
the relativistic models. It is also important to note that: 1) the
internal energy of the shocked ambient medium  takes the larger amount
of the injected energy in both cases ($\simeq 70\%$ of the injected
energy in the relativistic case; $\simeq 50\%$ in the classical case), and 2) the 
total amount of energy transferred to the ambient medium
(internal plus kinetic) is up to a $20\%$ larger in the relativistic
case at any time.
   
   A recent paper by \cite{wei17}, following a classical approach, but
including a contribution to pressure by cosmic rays and magnetic
field finds an increase in the efficiency of the heating of the
ambient medium. This result also shows that a contribution
different to kinetic energy (mainly in the form of heat-pressure)
to the amount of energy flux injected by the jet increases the
efficiency of heating. 

In another recent paper, \citet{ehk16} have performed a
series of simulations of relativistic (magnetized and unmagnetized)
jets covering a range in jet powers and velocities (from
non-relativistic, $v_j \simeq 0.25$c, to mildly-relativistic, $\Gamma_j
\simeq 3.2$). As discussed by the authors, the lack of bright hotspots
at the end of the lobes, in clear contrast with the observations of
powerful radio sources, tends to be alleviated by reducing the jet
cross section (unrealistically large in the original simulations). 
This result underpins the line of argument of this Letter.
 
Our results also suggest that the analytical
models that use the classical approach \citep[e.g.,][and all derived models]{ka97}, could underestimate the cavity/lobe
pressure created by relativistic jets, if the hotspot pressure is underestimated by using wide jet radii. 
Nevertheless, radiative models that deal with radio-source
brightness evolution and use the cavity pressure to compute it, either
assume jet parameters that imply small jet radii \citep[e.g.,][]{kda97,kb07}, or compute
$P_{h}$ from the observed size of hotspots \citep[e.g.,][]{bw07}, which alleviates this effect.

%
%

\subsection{Caveats}

The relation between the hotspot pressures in classical and
relativistic jets was obtained under the assumption that the terminal
shocks at the end of the jets are strong (see Sec.~\ref{mods}). This
is justified by the existence of the hotspots in powerful radio
sources that implies a substantial conversion of kinetic to internal
energy at the shock. Moreover, the resulting expression,
Eq.~(\ref{eq:pcpr}), was only used to connect both pressures along the
so-called 1D propagation phase of the relativistic jets, where the
assumption of the terminal shock being strong is more robust.
    
  We have also neglected the dynamical role of magnetic fields in 
this analysis. Magnetic field has been proposed to be slightly
below or close to equipartition (with the jet internal energy) along
the jet and at the radio lobes, which would make it dynamically
relevant at these scales \citep[see][and references
therein]{har15}. In case of equipartition, the magnetic
field represents a contribution to pressure in Eq.~(\ref{eq:L_rel})
\citep[see also, e.g.,][]{ehk16,wei17}. Therefore, we conclude that a
dynamical role of the magnetic field does not change the conclusions
derived in this work regarding the difference between the efficiency of
heating by relativistic jets, as compared to classical ones.
 
 \section{Conclusions}\label{conc}
  
We conclude  in this Letter the following key results:
\begin{itemize} 
\item The pressures in the hotspot and the cavity of the relativistic
  jets are always significantly  larger than in their classical
  counterparts. To sort out this discrepancy would require: i) hugely increase the jet rest-mass density with respect to that of
  relativistic jets, or ii) increase unrealistically the jet
  radius. Both solutions would imply huge mass-fluxes in conflict with
  current formation and acceleration mechanisms for jets, which
  propagate at relativistic speeds 
  at parsec-scales \citep[e.g.,][]{li09}. In the classical case, these mass fluxes would be $10^3-10^4$ times larger than those of a relativistic jet with $h_{j,r} = 2\,c^2 - 10\,c^2$, $\Gamma_{j,r} = 5 - 10$.  

\item{}The relativistic jets carve cavities whose volumes are 2 to 10
  times larger than the classical jets. This has a direct implication
  on the amount of heated gas. 
\item{}The heating efficiency of the relativistic jets is $\simeq
  20\%$ higher -- during the whole time evolution -- compared with the
  classical jets.        
\item{}As a consequence of previous results, the gas cooling rate in
  the cavity must strongly vary with position and time between both
  approaches, due to the different heating rates and energy deposition volumes.
  This should have a direct implication on the star formation rates of the host galaxies, and affect the subsequent
  galaxy evolution. 
\end{itemize}
 
To wrap this Letter up, we would like to stress that a relativistic approach is a crucial ingredient for
the study of radio mode feedback that cannot be obviated in a
consistent description of the galaxy evolution. 

\section*{Acknowledgements}
We thank the referee of the paper, G.V. Bicknell, for his constructive comments. 
This work has been supported by the Spanish Ministerio de Econom\'{\i}a y Competitividad
(grants AYA2013-40979-P, AYA2013-48226-C3-2-P, and AYA2016-77237-C3-3-P). 
MBLL thanks Spanish Ministerio de Educaci\'on  y
Cultura for a Collaboration Scholarship. 
Computer simulations have been carried out in 
Servei d'Inform\`atica de la Universitat de Val\`encia and in 
the Red Espa\~nola de Supercomputaci\'on (Tirant).






\bsp	
\label{lastpage}
\end{document}